\documentclass{book}
\usepackage{jfaa2e}
\usepackage{latexsym}		  
\usepackage{amsmath}
\usepackage{graphicx}
\newcommand{\edc}{\end{document}}
\newcommand{\Sff}{S_{f\!f}}
\newcommand{\domg}{\,d\Omega}
\newcommand{\bSff}{\mathbf{S}_{f\!f}}
\newcommand{\Sfg}{S_{f\!g}}
\newcommand{\Shh}{S_{hh}}
\newcommand{\Sfifi}{S_{\Phi\Phi}}
\newcommand{\bSfifi}{\mathbf{S}\mt_{\Phi\Phi}} 
\newcommand{\Sfiga}{S_{\Phi\Gamma}}
\newcommand{\mt}{^{(mt)}}
\newcommand{\kth}{^{(k)}}
\newcommand{\cov}{\mathrm{cov}}
\newcommand{\var}{\mathrm{var}}
\newcommand{\suml}{\sum\limits}
\newcommand{\bh}{\mathbf{h}}
\newcommand{\bB}{\mathbf{B}}
\newcommand{\bD}{\mathbf{D}}
\newcommand{\bF}{\mathbf{F}}
\newcommand{\bM}{\mathbf{M}}
\newcommand{\Lh}{L}

\issue{Volume XX, Issue XX, 2007}
\doi{jfaa---}
\begin{document}

\author{Mark A. Wieczorek and Frederik J. Simons}

\chapter[Minimum-variance multitaper spectral estimation on the
sphere]{Minimum-variance multitaper spectral estimation\\ on the
sphere}

\footnotetext{\textit{Math Subject Classifications.}
33C55, 34L05, 42B35, 42C10, 62M15.}

\footnotetext{\textit{Keywords and Phrases.}
Spherical harmonics, multitaper spectral analysis.}

\begin{abstract} 
We develop a method to estimate the power spectrum of a stochastic
process on the sphere from data of limited geographical coverage. Our
approach can be interpreted either as estimating the global power
spectrum of a stationary process when only a portion of the data are
available for analysis, or estimating the power spectrum from local
data under the assumption that the data are locally stationary in a
specified region. Restricting a global function to a spatial subdomain
--- whether by necessity or by design --- is a windowing operation,
and an equation like a convolution in the spectral domain relates the
expected value of the windowed power spectrum to the underlying global
power spectrum and the known power spectrum of the localization
window.  The best windows for the purpose of localized spectral
analysis have their energy concentrated in the region of interest
while possessing the smallest effective bandwidth as possible. Solving
an optimization problem in the sense of Slepian (1960) yields a family
of orthogonal windows of diminishing spatiospectral localization, the
best concentrated of which we propose to use to form a weighted
multitaper spectrum estimate in the sense of Thomson (1982). Such an
estimate is both more representative of the target region and reduces
the estimation variance when compared to estimates formed by any
single bandlimited window. We describe how the weights applied to the
individual spectral estimates in forming the multitaper estimate can
be chosen such that the variance of the estimate is minimized.

\end{abstract}

\section{Introduction}

Spectral analysis is an indispensable tool in many branches of the
physical and mathematical sciences, with common applications ranging
from one-dimensional time series to two-dimensional image
analysis. For many purposes it is sufficient to employ a Cartesian
geometry, and for this case a plethora of sophisticated techniques
have been developed, such as parametric, maximum-entropy, and
multitaper spectral analysis (see~\cite{Percival+93} for a
comprehensive review). However, for certain problems, especially in
geophysics, it is necessary to obtain spectral
estimates from data that are localized to specific regions on the
surface of a sphere. While subsets of these data could be mapped to a
two-dimensional plane, enabling the use of Cartesian methods, this
procedure is bound to introduce some error into the obtained spectral
estimates. As the size of the region approaches a significant fraction
of the surface area of the sphere, these would naturally become
increasingly unreliable.

Spectral analysis on the sphere is an important tool in several
scientific disciplines, and two examples suffice to illustrate the
range of problems that are often encountered. First, in geophysics and
geodesy it is common to represent the gravity field and topography of
the terrestrial planets as spherical harmonic expansions and to use
their cross-spectral properties to investigate the interior structure
of the body~\cite{Wieczorek2007}. However, since the relationship
between the gravity and topography coefficients depends upon the
geologic history of the geographic area of interest, it is often
necessary to consider only a localized subset of these data. A second
example is in the field of cosmology where the power spectrum of the
cosmic microwave background radiation is used to place important
constraints on the structure and constitution of the universe
\cite{Spergel+2006}. In contrast to many geophysical problems, the
measured temperature fluctuations are often assumed to be derived from
a globally stationary process. Nevertheless, when estimating the power
spectrum from satellite and terrestrial-based measurements, it is
necessary to mask out regions that are contaminated by emissions
emanating from the plane of our own galaxy~\cite{Hinshaw+2006}.

On the sphere one is thus concerned generally with estimating the
power spectrum of a certain process from data confined to a restricted
region. As illustrated by the above examples, this can be interpreted
in one of two ways. In one case, the function is assumed to be
stationary, and an estimate of the global power spectrum is desired
based on a localized subset of data. In the second case, the data are
known to be non-stationary, and an estimate of a ``localized'' power
spectrum is sought by assuming local stationarity of the data within
the specified region.

The restriction of data to a specified region is equivalent to
multiplying a globally defined function by a localization window or
``data taper'', and the objective is to relate the power spectrum of
this localized field to the global one. This has been investigated by
using single binary masks~\cite{Hivon+2002}, as well as families of
orthogonal isotropic windows with the purpose of obtaining a
``multitaper'' estimate~\cite{Wieczorek+2005}. The use of multiple
localization windows, as originally pioneered in the Cartesian domain
by Thomson~\cite{Thomson82}, possesses many advantages over that of a
single window, in particular, smaller variances of the resulting
spectral estimates and more uniform coverage of the localization
region. Here, we extend our previous approach~\cite{Wieczorek+2005} of
using zonal tapers (i.e., those with azimuthal symmetry about a polar
axis) to the general case that includes non-zonal
tapers~\cite{Simons+2006}. The inclusion of non-zonal data tapers
greatly increases the number of individual spectral estimates that
make up the multitaper estimate, and this leads to a significant
reduction in the variance of this estimate. We
demonstrate how the weights of the individual estimates can be chosen
to minimize the multitaper estimation variance.

In this paper, we first describe the theory of estimating the global
power spectrum of a stochastic stationary process through the use of
localization windows. This includes quantifying the relationship
between the global and localized power spectra, the design of windows
that are optimal for this purpose, and the formation of a
minimum-variance multitaper spectral estimate. The majority of the
theoretical development relating to these problems is contained in
four appendices. Following this, we describe the statistical
properties of the multitaper estimates (such as their bias and
variance) heuristically when the data are governed by a stochastic
process with either a ``white'' or ``red'' power spectrum. Next, we
give an example of estimating the power spectrum from a single
realization of a stochastic process. Finally, we conclude by
emphasizing avenues of future research.

\section{Theory}

\subsection{Localized spectral estimation}
\label{localization}

Any real square-integrable function on the unit sphere can be expressed by
a linear combination of orthogonal functions as 

\begin{equation}
f(\Omega) = \sum_{l=0}^{\infty} \sum_{m=-l}^{l} f_{lm} Y_{lm}(\Omega)
,
\label{eq:sh_ex1}
\end{equation} 
where $Y_{lm}$ is a real spherical harmonic of degree~$l$ and order
$m$, $f_{lm}$ is the corresponding expansion coefficient, and
$\Omega=(\theta,\phi)$ represents position on the sphere in terms of
colatitude, $\theta$, and longitude, $\phi$. The real spherical
harmonics are defined in terms of a product of a Legendre function in
colatitude and either a sine or cosine function in longitude,

\begin{equation}
\label{eq:sh_def}
Y_{lm}(\Omega) = \left \lbrace \begin{array}{ll} \bar{P}_{lm}(\cos
	\theta) \cos m \phi & \mbox{if $m \ge 0$} \\
	\bar{P}_{l|m|}(\cos \theta) \sin |m| \phi & \mbox{if $m < 0$},
	\end{array} \right.
\end{equation}
and the normalized Legendre functions used in this
investigation are given~by 
\begin{equation}
\label{eq:legendre_norm}
\bar{P}_{lm}(\mu)=\sqrt{(2-\delta_{0m}) (2l+1)}\sqrt{
\frac{(l-m)!}{(l+m)!}}\, P_{lm}(\mu)
,
\end{equation}
where $\delta_{ij}$ is the Kronecker delta function and $P_{lm}$ is
the standard associated Legendre function,

\begin{equation}
\label{eq:Aplm}
P_{lm}(\mu) =  \frac{1}{2^l l!} \left( 1-\mu^2\right)^{m/2}\left(
\frac{d}{d\mu}\right)^{l+m} \left(\mu^2-1\right)^l 
.
\end{equation}
With the above definitions, the spherical harmonics of~(\ref{eq:sh_def})
are orthogonal over the sphere and possess unit power,
\begin{equation}
\label{eq:sh_orthog}
\frac{1}{4\pi} \int_\Omega Y_{lm}(\Omega) Y_{l'm'}(\Omega) \domg = 
\delta_{ll'}\, \delta_{mm'}
,
\end{equation}
where $d\Omega = \sin \theta \, d\theta \, d\phi$. We note that this
unit-power normalization is consistent with that used by the geodesy
community, but differs from the physics and seismology communities
that use orthonormal harmonics~\cite{Dahlen+98,Varshalovich+88}
and the magnetics community that uses Schmidt semi-normalized
harmonics~\cite{Blakely95}.  Furthermore, we omit the Condon-Shortley
phase factor of $(-1)^m$ in the definition of the spherical harmonics,
which is consistent with the usage of the geodesy and magnetics
communities. Using~(\ref{eq:sh_ex1}) and~(\ref{eq:sh_orthog}), the
total power of a real function $f$ can be shown to be related to its
spectral coefficients by a generalization of Parseval's theorem:
\begin{equation}
\label{eq:power}
\frac{1}{4\pi} \int_\Omega \left[ f(\Omega)\right]^2\domg =
\sum_{l=0}^\infty \Sff(l) 
,
\end{equation}
where
\begin{equation}
\label{eq:power_spectrum}
\Sff(l) = \sum_{m=-l}^l  f_{lm}^2 
\end{equation}
is referred to as the power spectrum of $f$. Similarly, the
cross-power of two real functions $f$ and $g$ is given by 
\begin{equation}
\frac{1}{4\pi} \int_\Omega f(\Omega) \,g(\Omega) \domg =
\sum_{l=0}^{\infty} \Sfg(l) , 
\end{equation}
where the cross-power spectrum is
\begin{equation}
\Sfg(l) = \sum_{m=-l}^l f_{lm} \, g_{lm}
.
\end{equation}
The power and cross-power spectra possess the property that they are
unmodified by a rotation of the coordinate system
\cite{Kaula67a}. Efficient and accurate algorithms for calculating the
normalized Legendre functions and spherical harmonic coefficients of a
function can be found in~\cite{Driscoll+94} and~\cite{Holmes+2002},
respectively. The relationship between real and complex spherical
harmonics, which is necessary for certain derivations in the
appendices, is presented in Appendix A.

If the function $f$ of~(\ref{eq:sh_ex1}) were known globally, it would
be a trivial matter to obtain its spherical harmonic coefficients
$f_{lm}$, and by consequence $\Sff$, its power
spectrum~\cite{Dahlen+2008}. Unfortunately, in spherical analyses it
is common that the function is only known within a restricted domain
on the sphere, or conversely, that it is necessary to ignore certain
contaminated or unrepresentative regions of the global data
set. Alternatively, it may arise that the global function is known to
be non-stationary, and that a local power spectrum estimate is desired
under the assumption of local stationarity for a specified region.

The localization of a global function to a given domain can be
formulated as a windowing operation, 
\begin{equation}
\Phi(\Omega) = h(\Omega) \, f(\Omega),
\end{equation}
where $h$ is the localization window and $\Phi$ is the localized
version of $f$. While a naive binary mask is sometimes used for $h$,
indicating either the presence or absence of data, this choice will
later be shown to possess undesirable properties analogous to those of
the standard periodogram~\cite{Percival+93}. After having chosen~$h$,
the power spectrum of the localized function $\Sfifi$ is easily
calculated from~(\ref{eq:power}) and~(\ref{eq:power_spectrum}). It
should be clear that while $\Sfifi$ might resemble $\Sff$, the two
will not be equal as a result of the windowing operation. The quantity
$\Sfifi$ will here be referred to as both the \emph{localized} and
\emph{windowed} power spectrum.

The relationship between $\Sff$ and $\Sfifi$ is a complicated one, and
it is generally not possible to invert for the former given the
latter. Nevertheless, if it is assumed that the function $f$ is a
stationary stochastic process, a simple relationship exists between
the \emph{expectation} of $\Sfifi$ and $\Sff$, namely,
\begin{equation}
\left\langle \Sfifi(l) \right\rangle = 
 \suml_{j=0}^{L} \Shh(j)
 \suml_{i=|l-j|}^{l+j} \Sff(i)  \left(C_{j0i0}^{l0}\right)^2,
 \label{eq:bias}
\end{equation}
where $\langle\cdots\rangle$ denotes the expectation operator, $L$ is
the spherical harmonic bandwidth of $h$, and the symbol in parentheses
is a Clebsch-Gordan coefficient. Various forms of this relationship
have been previously derived independently
\cite{Hivon+2002,Peebles73,Wieczorek+2005}, and in Appendix B we
generalize this to the case of localized cross-power spectra. Here, it
is sufficient to note that for~(\ref{eq:bias}) to hold the spherical
harmonic coefficients of $f_{lm}$ are required to be zero-mean random
variables with a variance that depends only on degree~$l$. The
expectation of the localized power spectrum is to be considered as an
average over all possible realizations of the random variables
$f_{lm}$.

Equation~(\ref{eq:bias}) shows that the expectation of the localized
power spectrum is related to the power spectrum of $f$ and the
localization window $h$ by an operation reminiscent of a
convolution. In particular, it is important to note that each
degree~$l$ of the localized power spectrum contains contributions from
the global spectrum $\Sff$ within the degree range $l\pm L$. Thus, the
spherical harmonic bandwidth of the localization window directly
controls how the global power spectrum $\Sff$ is ``smoothed'' in
determining the expectation of the localized spectrum $\Sfifi$. Given
the power spectrum of a localization window, as well as an estimate of
the localized power spectrum expectation, as discussed in
Section~\ref{section:single} and~\cite{Wieczorek+2005}, several
techniques could be used to invert for the global spectrum. This would
make the resulting ``deconvolved'' localized spectral estimate
statistically unbiased, but would come at the cost of a higher
estimation variance \cite{Dahlen+2008}.

\subsection{Window design}
\label{window}

The convolution-type operation of~(\ref{eq:bias}) that relates the
localized power spectrum to the global and localization window power
spectra demonstrates the importance of using a window $h$ with as
small a spherical harmonic bandwidth as possible. For instance, if a
naive binary mask were used to isolate certain domains on the sphere,
the bandwidth of this window would be infinite, and every localized
spectral estimate $\Sfifi$ would be influenced by every degree of the
global spectrum $\Sff$. If the global power spectrum possessed a
significant dynamic range, degrees with high power could positively
bias the localized power spectrum at degrees where the global power
spectrum is small. Such spectral leakage could hinder attempts to
invert for the global spectrum $\Sff$ given knowledge of~$\Shh$
and~$\Sfifi$.

In order to spatially localize a function on the sphere, it is clear
that the localization window should possess zero or near-zero
amplitudes exterior to a specified region of interest
$R$. Additionally, in order to limit the effects of spectral leakage,
the effective spectral bandwidth of the window should be as small as
possible. The problem of designing windows that are ideally localized
in both the space and spectral domains was originally posed as an
optimization problem by Slepian and coworkers in the Cartesian domain
(see~\cite{Slepian83} for a review), and later extended to the sphere
by various authors
\cite{Grunbaum+82,Simons+2006b,Simons+2006,Wieczorek+2005}.

One form of this optimization problem is to find those functions whose
spatial power is concentrated within a given domain $R$, but yet are
bandlimited to a spherical harmonic degree~$L$; i.e., to find those
functions that maximize the ratio
\begin{equation}
\label{eq:concenration}
\lambda = \int_R h^2(\Omega)\domg \,  {\bigg /} \int_{\Omega}
h^2(\Omega) \domg 
.
\end{equation}
It can be shown~\cite{Simons+2006,Wieczorek+2005} that
this equation reduces to an eigenvalue equation
\begin{equation}
\bD \, \bh = \lambda \, \bh
,
\label{eq:eig1}
\end{equation}
where $\bh$ is a vector of length $(L+1)^2$ containing the
spherical harmonic coefficients of the window, and $\bD$ is a $(L+1)^2
\times (L+1)^2$ localization kernel. The solution
of this eigenvalue problem yields a family of orthogonal windows
$h\kth$ (which we normalize to have unit power), with corresponding
spatial concentration factors ordered such that
\begin{equation}\label{evals}
1 > \lambda_1 \ge  \cdots \lambda_k \cdots \ge \lambda_{(L+1)^2} > 0.
\end{equation} 
As the eigenvalue spectrum has been found empirically to transition
quickly from values near unity to zero, the sum of the eigenvalues
corresponds approximately to the number of windows with good
spatiospectral localization properties, and this ``Shannon number''
has been shown~\cite{Simons+2006} to be equal to
\begin{equation}
N= \sum_{k=1}^{(L+1)^2} \lambda_k = (L+1)^2 \frac{A}{4 \pi},
\end{equation}
where $A$ is the area spanned by the region $R$ on the unit
sphere.

An alternative criterion for designing a localization window is to
instead find those functions that are perfectly contained within a
region $R$ and whose spectral power is concentrated within a spherical
harmonic bandwidth $L$; i.e., to maximize
\begin{equation}
\lambda = \sum_{l=0}^{L} \sum_{m=-l}^l h_{lm}^2\,  {\bigg /}
\sum_{l=0}^{\infty} \sum_{m=-l}^l h_{lm}^2, 
\end{equation} 
where the function $h$ is defined to be zero exterior to $R$. This
spectral optimization problem is complementary to the spatial
concentration problem~\cite{Simons+2006,Wieczorek+2005}. In
particular, the functions are identical within the domain $R$, the
spectral- and spatial-concentration eigenvalue spectra are identical
up to the $(L+1)^2$-th eigenvalue, and the power spectra of the
functions for $l\le L$ differ only by a factor equal to the square of
the eigenvalue. Thus, if one desires localization windows that are
perfectly restricted to $R$, yet nonetheless possess some spectral
power beyond the effective bandwidth $L$, it is only necessary to
compute the space-concentration windows by means of~(\ref{eq:eig1})
and to then set these functions equal to zero exterior to $R$.

The geometry of the localization domain $R$ determines the form of the
localization kernel $\bD$. While this matrix and its eigenvalues and
eigenfunctions can be computed for any arbitrary
domain~\cite{Simons+2006}, certain geometries greatly simplify this
task. For bandlimited windows, if the domain $R$ is a spherical cap
located at the North pole ($\theta=0$), $\bD$ is block diagonal,
allowing the eigenvalue problem of~(\ref{eq:eig1}) to be solved separately for
each individual angular order $m$. More importantly, there exists a
tridiagonal matrix with analytically prescribed elements that commutes
with~$\bD$, and hence shares the same
eigenfunctions~\cite{Grunbaum+82,Simons+2006}.  Similarly, for the
case of two antipodal spherical caps (or equivalently, an equatorial
belt), a simple commuting tridiagonal matrix exists as
well~\cite{Simons+2006b}.

\begin{figure}[b]
\centering
{\includegraphics[width=\columnwidth]{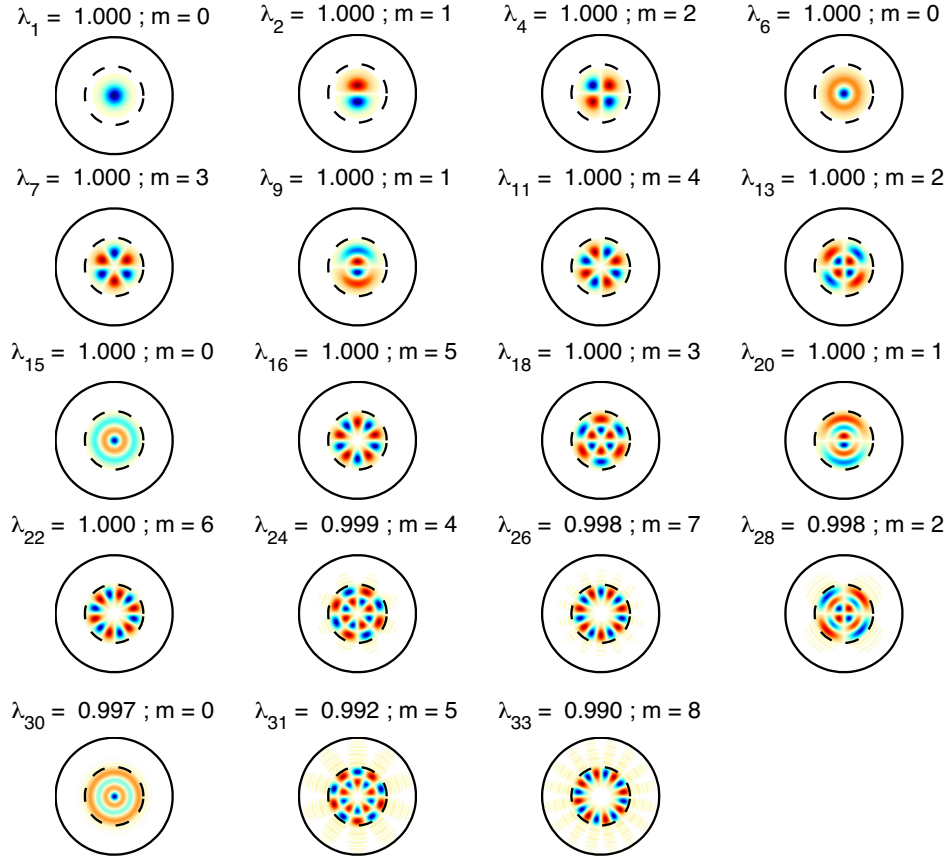}}
\caption{Spatial rendition of those functions with angular order
$m\ge0$ that are nearly perfectly concentrated within a spherical cap
of angular radius $\theta_0=30^\circ$ and with a spherical harmonic
bandwidth of $L=29$. For this case, $N\simeq60$, and the total number
of windows with $\lambda>0.99$ is~34. Each non-zonal function possess
a twin with angular order $-m$ that is rotated azimuthally by
$90^\circ/m$ (not shown).  \label{fig:tapers}}
\end{figure}

The theoretical development presented in this paper is valid for any
irregularly shaped concentration domain, and for either space-limited
or bandlimited localization windows. In this investigation, however,
we will employ exclusively bandlimited localization windows whose
spatial power is optimally concentrated within a spherical cap located
at $\theta=0^\circ$.  A spherical-cap concentration domain is a
close analog to the square domain that is commonly used in 2-D
Cartesian analyses and is likely to find broad applicability to many
problems \cite{Belleguic+2005,Han+2007,McGovern+2002,Wieczorek2007b}.

For demonstration purposes, we use a concentration domain with an
angular radius of $\theta_0=30^\circ$ and a spectral bandwidth of
$L=29$, corresponding to $N\simeq60$. We will further restrict
ourselves to those windows that are nearly perfectly localized with
$\lambda>0.99$, of which there are 34. As illustrated in
Figure~\ref{fig:tapers}, each of these windows possess non-zero spherical harmonic
coefficients for only a single angular order $m$. Whereas only windows
with $m\ge 0$ are shown in this figure, we note that each non-zonal
window (i.e, $m\ne 0$) has an identically concentrated twin of angular
order $-m$ that differs only by an azimuthal rotation of
$90^\circ/m$. In contrast, if we were to employ only those windows
that are isotropic (i.e., $m=0$), we would have only~4 nearly
perfectly concentrated windows at our disposal. In that case, as shown
by~\cite{Wieczorek+2005}, the zonal Shannon number would be given by
$N_0=(L+1)\, \theta_0/ \pi=5$.

\subsection{Minimum-variance multitaper spectral estimation}
\label{minvar}

Several studies have used single localization windows to obtain
localized spectral estimates on the sphere
\cite{Belleguic+2005,Hivon+2002,McGovern+2002,Simons+97a}.
Nevertheless, as originally formulated by Thomson~\cite{Thomson82},
the use of multiple orthogonal localization windows can have
significant advantages over the use of any single window
\cite{Percival+93,Walden+94,Wieczorek+2005}. In particular, the energy
of a single bandlimited window will always non-uniformly cover the
desired concentration region, and this will result in some data being
statistically over- or under-represented when forming the spectral
estimate. In contrast, the cumulative energy of the orthogonal windows
solving~(\ref{eq:eig1}) more uniformly covers the concentration
region. Second, since the spectral estimates that result from using
orthogonal windows are somewhat uncorrelated, a multitaper average of
these will possess a smaller estimation variance. This is especially
important since most investigations are limited to analyzing a single
realization of a stochastic process.

\begin{figure}[b]
\centering
{\includegraphics{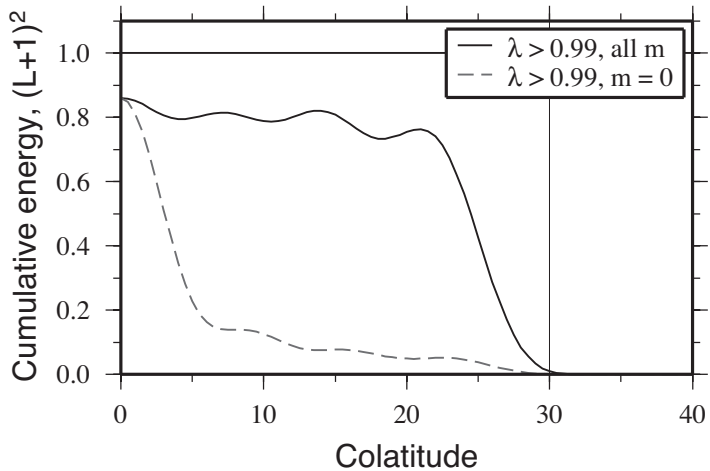}}
\caption{Cumulative energy of the nearly perfectly concentrated
($\lambda>0.99$) localization windows of Figure~\ref{fig:tapers}
($\theta_0=30^\circ$ and $L=29$). The solid curve is the cumulative
energy for all 34 tapers, whereas the dashed curve is for the subset
of the four zonal tapers. If all $(L+1)^2$ tapers were employed, the
cumulative energy would be $(L+1)^2$ everywhere. \label{fig:cum_energy}}
\end{figure}

To demonstrate the first of the above advantages of using multiple
orthogonal localization windows, we plot the cumulative energy (i.e.,
the squared amplitude) of the windows utilized in this study as a
function of colatitude in Figure~\ref{fig:cum_energy}. Since windows
of angular order $m$ and $-m$ are included, this function is
independent of azimuth $\phi$. In particular, the cumulative energy of
the four zonal windows is shown, as well as that for the 34 zonal and
non-zonal windows of Figure~\ref{fig:tapers}. As is readily seen, the
energy of the four zonal windows is peaked near the center of the
spherical cap, giving lesser importance to data located close to the
cap edge. In contrast, the cumulative energy of all zonal and
non-zonal windows is more evenly spread across the concentration
domain. While data adjacent to the cap edge are still somewhat
downweighted, this is not nearly as drastic as for the case when only
zonal tapers are used.  We note that if all $(L+1)^2$ localization
windows were to be used, the cumulative energy would be $(L+1)^2$~
everywhere~\cite{Simons+2006}, though in this case, the resulting
multitaper estimate would not be localized to any particular spatial
region of the sphere.

We define the multitaper localized power spectrum estimate of a
function $f$ as a weighted average of direct spectral estimates
obtained from $K$ orthogonal tapers:
\begin{equation}
S\mt_{\Phi \Phi}(l) = \sum_{k=1}^K a_k \,
S\kth_{\Phi \Phi}(l)
\label{eq:mt2}
,
\end{equation}
with the constraint that the sum of the weights is unity,
\begin{equation}
\suml_{k=1}^K a_{k}=1. 
\label{eq:mt2a}
\end{equation}
It is clear from~(\ref{eq:bias}) and~(\ref{eq:mt2}) that the
expectation of this estimate is given by
\begin{equation} 
\left\langle \Sfifi\mt(l) \right\rangle = 
 \suml_{j=0}^{L} \left( \suml_{k=1}^K a_k \,\Shh\kth(j) \right)
 \suml_{i=|l-j|}^{l+j} \Sff(i)  \left(C_{j0i0}^{l0}\right)^2,
 \label{eq:biasmt}
\end{equation} 
and that its bias is 
\begin{equation}
\mathrm{bias}\left\{ \Sfifi\mt(l) \right\}= \suml_{k=1}^K a_k
\left(\left\langle \Sfifi\kth(l) \right\rangle - \Sff(l)\right) =
\suml_{k=1}^K a_k \, B_{k} ,
\label{biasdef}
\end{equation}
where the elements of the vector $\bB$ depend implicitly upon the
power spectrum of the global function, the power spectrum of the k-th
localization window, and the spherical harmonic degree~$l$.  Using the
above definitions, we show in Appendix C that the variance of the
multitaper estimate can be written as
\begin{equation} 
\label{eq:mt_varfinal}
\var\! \left\{S\mt_{\Phi \Phi}(l)
\right\}=  \sum_{j=1}^K \sum_{k=1}^K a_j \,  F_{jk}\,a_k,
\end{equation} 
where the symmetric covariance matrix $\bF$ depends both upon
$\Sff(l)$ and the expansion coefficients of the window, and is given
by the lengthy equations~(\ref{eq:f}) and~(\ref{eq:fjk2}). In Appendix~C
we generalize the expression~(\ref{eq:mt_varfinal}) to apply to
cross-power spectra as well. Appendix D describes how to calculate the 
covariance of two multitaper spectral estimates at two different degrees.

We next address the question of which values to use for the weights
$a_j$ when constructing a multitaper estimate. Two cases that are in
common use in the time series community are either to take weights
that are all equal to $1/K$, or weights that are proportional to the
eigenvalues~(\ref{evals}) of the localization windows (this latter
case helps simplify some mathematical
relationships~\cite{Dahlen+2008,Wieczorek+2005}). An alternative
approach would be to instead solve for those weights that minimize
some combination of the variance and bias of the estimate. As an
example, if it were important to obtain estimates that possessed both
low variance and low bias, then an appropriate measure to minimize
might be the mean-squared error of the spectral estimate at a
particular degree~$l$, which is simply a sum of the variance and
squared bias: 
\begin{equation}\label{mse}
\mathrm{mse}=\var +\mathrm{bias}^2=
\sum_{j=1}^K \sum_{k=1}^K a_j \, \left(F_{jk} + B_jB_k \right) \, a_k
.
\end{equation}
Thomson~\cite{Thomson82} advocated a similar (though approximate)
approach that he referred to as ``adaptive weighting.'' However, since
his study used windows that were not bandlimited, the choice was made
to consider only that portion of the bias that resulted from
frequencies greater than the effective bandwidth of the window (i.e.,
the broadband bias).  Since both the covariance matrix $\bF$ and bias
$\bB$ depend upon the unknown global power spectrum, it is clear that
such a minimization procedure would, in general, be iterative.

Instead of attempting to minimize the mean-squared error, a different
philosophy is to solve for those weights $a_j$ that minimize solely
the variance of the multitaper spectral estimate. While the bias of
such an estimate would naturally be larger, this potentially
unfavorable characteristic is countered by the fact that the bias is
completely \emph{quantifiable} when the global power spectrum is
known. For a large class of inverse problems, one is concerned with
comparing the biased multitaper spectral estimate directly to a
similarly biased theoretical model. In this situation, all that is
important is how the goodness-of-fit between the two spectra varies as
a function of the theoretical model parameters, and not how closely
the windowed power spectra match their global equivalents. When it is
easy to account for the estimation bias, the relevant quantity to
minimize is naturally the variance of the windowed spectral estimates.
Since many scientific problems that use multitaper analyses are done
so in the context of comparing forward models to the observations,
minimum-variance multitaper spectral estimation will be emphasized in
the following sections.

The numerical values of the weights $a_j$ that minimize the
mean-squared error of (\ref{mse}) will here be solved for subject to
the constraint that the sum of the weights is unity. (To obtain the
minimum variance solution, it is only necessary to set the vector
$\bB$ to zero.) This is easily accomplished by minimizing the objective
function
\begin{equation}
\Psi = \sum_{i=1}^K \sum_{j=1}^K a_i \, \left(F_{ij} + B_iB_j \right)
\, a_j + \lambda \left( \sum_{k=1}^K a_k -1\right)
\end{equation} 
with respect to the weights $a_j$ and Lagrange multiplier $\lambda$,
which yields the following set of linear equations:
\begin{eqnarray}
\frac{\partial\Psi}{\partial a_n} & = & \suml_{i=1}^K a_i \,
\left(F_{in}+B_iB_n\right) + \suml_{j=1}^K a_j \,
\left(F_{nj}+B_nB_j\right)+ \lambda = 0, \\
\frac{\partial\Psi}{\partial\lambda} &=& \sum_{k=1}^K a_k -1 = 0.
\end{eqnarray} 
Since $\bF$ is symmetric, these equations
can be written in matrix notation as 
\begin{equation}
\left[ \begin{array}{cccc} 2 \left(F_{11}+B_1B_1\right) & \cdots & 2
\left(F_{1K}+B_1B_K\right) & 1 \\ 2 \left(F_{21}+B_2B_1\right) &
\cdots & 2 \left(F_{2K}+B_2B_K\right) & 1 \\ \vdots & \ddots & \vdots
& \vdots \\ 2\left( F_{K1}+B_KB_1\right) & \cdots & 2
\left(F_{KK}+B_KB_K\right) & 1 \\ 1 & \cdots & 1 & 0 \\
\end{array}\right]
\left[ \begin{array}{c}
a_1 \\ a_2 \\\vdots \\ a_K \\ \lambda \end{array} \right] 
= \left[  \begin{array}{c}
0 \\ 0 \\  \vdots \\ 0 \\ 1 \end{array}\right].
\label{eq:minvar}
\end{equation} 
In order to find the optimal values of the weights $a_j$ (as well as
$\lambda$, which is not further needed), it is only necessary to solve
a simple linear equation. If one needs to calculate the covariance
matrix $\bF$, the weights are easily obtained with little additional
computational effort.

\section{White and Red Stochastic Processes}
\label{whitered}

Many physical processes obey power-law behavior in the sense that
their power spectrum varies as
\begin{equation} 
\label{eq:powerlaw}
\Sff (l)\sim l^{\beta}.
\end{equation} 
When the exponent $\beta$ is equal to zero, the total power per
spherical harmonic degree is constant, and we refer to the process as
\emph{white}. In contrast, when the exponent is less than zero, the power
decreases with increasing spherical harmonic degree and the spectrum
is \emph{red}. We note that this terminology depends on the definition of
the power spectrum~(\ref{eq:power_spectrum}), which may differ between
fields of application~\cite{Dahlen+2008}. Common examples of red
spectra include planetary gravitational fields and topography. In this
section, we take two representative values of $\beta$, namely $0$ and
$-2$, and describe in detail the properties of localized power
spectrum estimates using the windows described in
Section~\ref{window}. For a discussion on Cartesian spectral analysis
of power law processes, see~\cite{McCoy+98}.

\begin{figure}[b]
{\includegraphics[width=\columnwidth]{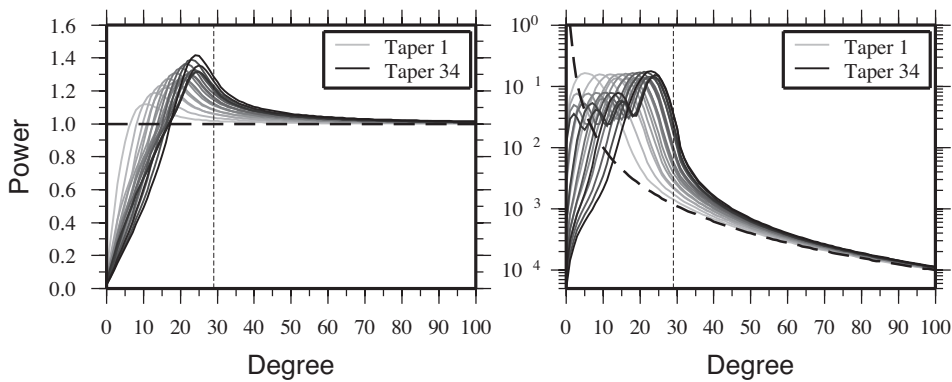}}
\caption{Expectations of localized power spectra for stationary
stochastic global processes. Heavy dashed lines represent white (left)
and red (right) global power spectra with $\beta=0$ and $-2$,
respectively. The expectations of the localized spectra were generated
using each of the 34 localization windows shown in
Figure~\ref{fig:tapers}; the taper numbers increase from light gray to
black. Vertical lines denote the spectral bandwidth $L=29$ of the
tapers. \label{fig:bias}}
\end{figure}

The relationship between a global power spectrum $\Sff$ and the
expectation of its localized equivalent $\Sfifi$ is described
by~(\ref{eq:bias}). In particular, for a given degree~$l$, the
expectation of the localized spectrum depends upon the global power
spectrum within the degree range $l\pm L$, where $L$ is the bandwidth
of the localization window. Thus, while $\Sfifi$ should be expected to
resemble the global spectrum, it will nevertheless be biased. In Figure~\ref{fig:bias}, we plot the localized power
spectra of a white (left) and red (right) power law process using the
34 windows displayed in Figure~\ref{fig:tapers}. As is readily seen,
the localized spectra do indeed resemble the global spectra (shown by
the heavy dashed lines), and appear to asymptotically approach the
global values at high degrees. However, for degrees close to or less
than the bandwidth of the window, the bias can be appreciable. In
particular, for degrees close to zero the localized power spectrum is
always biased down, whereas for degrees greater than about $L/2$ the
bias is positive.

\begin{figure}[b]
{\includegraphics[width=\columnwidth]{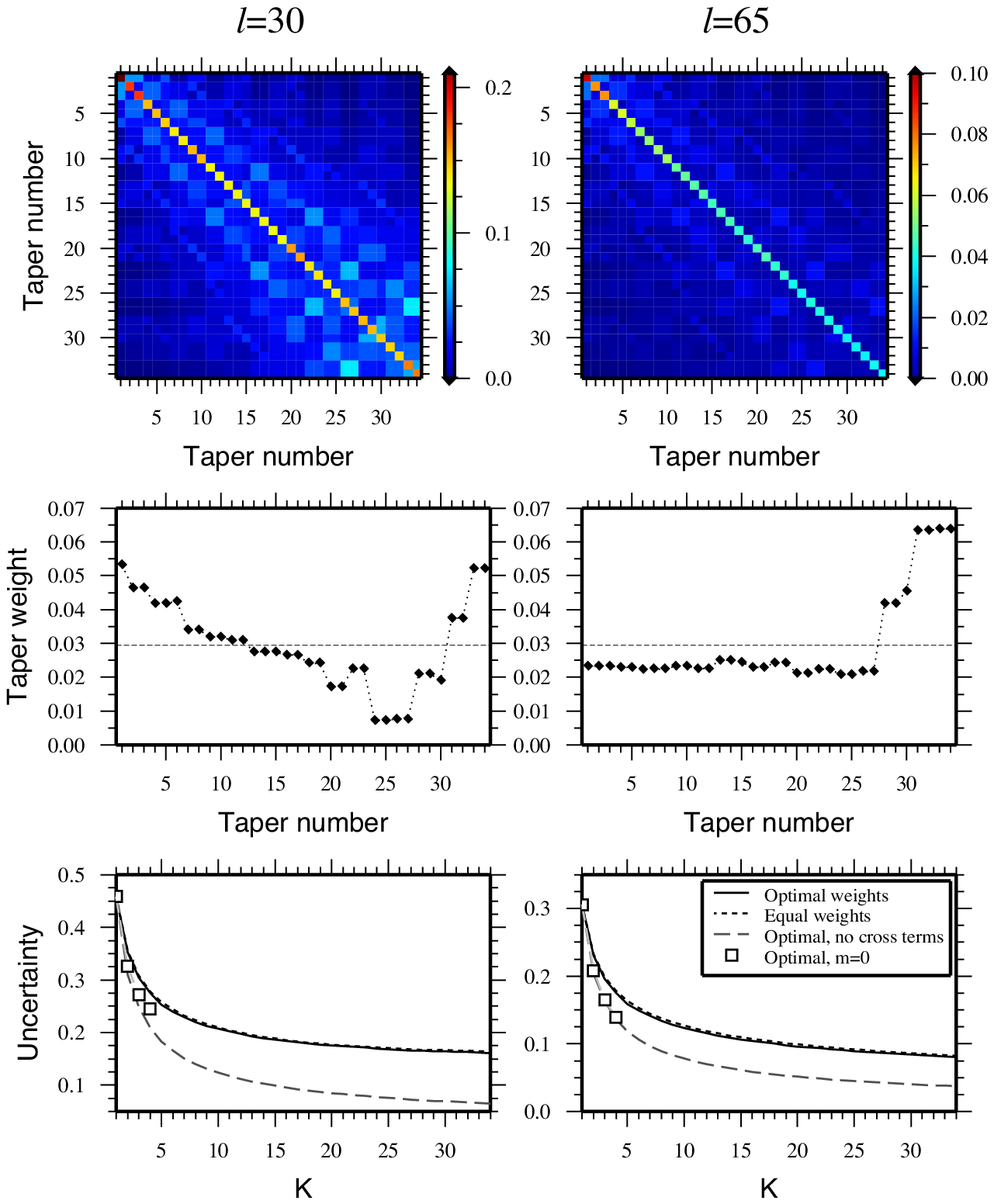}}
\caption{(top) Covariance matrix between the 34 best concentrated
tapers of the spherical-cap concentration problem in
Figure~\ref{fig:tapers} for a white global spectrum at $l=30$ (left)
and $65$ (right). (center) Optimal weights, $a_j$, that minimize the
variance of the spectral estimate at $l=30$ and $65$ when all 34
tapers are used. The dotted horizontal line corresponds to equal
weights of $1/34$. (bottom) Uncertainty of the spectral estimate at
$l=30$ and $65$ as a function of the number of employed tapers. Shown
are the cases of using optimal weights, equal weights, optimal weights
if all the off-diagonal terms of the covariance matrix were zero, and
optimal weights for only the zonal tapers. \label{fig:c_0_30}}
\end{figure}

The bias is clearly worse for the red-spectrum example than for the
white example. This is simply a result of the fact that for red
spectra, the range of the global power spectrum within the degree
range $l\pm L$ is greater for small $l$ than for large $l$. As a
result, the high power at small $l$ in the global spectrum will
disproportionately influence the sum in~(\ref{eq:bias}), ``leak''
towards higher degrees, and bias the localized power spectrum
upwards. In contrast, for large $l$, the global power spectrum can be
considered to be approximately constant within the degree range that
contributes to the localized spectral estimate, and the bias
properties are similar to those in the white-spectrum example. We note
that for degrees greater than the bandwidth of the window, the bias
appears to increase with increasing taper number. This is simply
related to the shape of the power spectrum of the localization windows
\cite{Simons+2006,Wieczorek+2005}. For the best localization windows,
the power is concentrated at the lowest degrees, which acts to give
the window an ``effective'' bandwidth that is somewhat less than
$L$. In contrast, for higher taper numbers, the power becomes more
evenly spread across the nominal bandwidth, ensuring that all degrees
in the range $l\pm L$ contribute appreciably to the sum
in~(\ref{eq:bias}).

The variance properties of localized multitaper spectral estimates at
degrees $l=30$ (left) and $65$ (right) are displayed in
Figure~\ref{fig:c_0_30} for the case where the global power spectrum
is white. The upper panel plots the covariance matrix~$\bF$
(\ref{eq:mt_varfinal}), and as is seen, the largest contributors are
the diagonal terms, indicating that the spectral estimates from
individual windows are not too highly correlated. The off-diagonal
terms appear to be relatively less important for the case of $l=65$ in
comparison to $l=30$, and this trend continues with increasing
degree. Thus, as quantified below, the individual spectral estimates
that make up the multitaper estimate can be considered as being
somewhat statistically independent, with the level of this
approximation improving with increasing degree.

The middle panel of Figure~\ref{fig:c_0_30} shows the weights that
minimize the variance of the localized multitaper spectral estimate
when all 34 tapers are used, and the lower panel shows its square
root, the uncertainty, as a function of $K$. Somewhat surprisingly,
even though the optimal weights~$a_j$ are decidedly non-uniform, the
uncertainty of the multitaper estimate using these weights (solid
curves) differs only insignificantly from what would arise if the
weights were all equal (short-dashed curves). For comparative
purposes, we plot the optimal uncertainty using only the four zonal
localization windows (squares). As noted by~\cite{Wieczorek+2005}, the
multitaper estimates using zonal windows are nearly uncorrelated, and
their uncertainty decreases as $1/\sqrt K$. By including the 30
non-zonal windows, which are equally well concentrated as the zonal
ones, the uncertainty has been reduced by an additional 50\%. The
uncertainty that would arise if all the individual spectral estimates
were completely uncorrelated is also shown, calculated by setting the
cross-terms of the covariance matrix equal to zero (long-dashed
curves). This variance is lower in magnitude than the corresponding
curve that includes the cross terms, which demonstrates that while the
non-zonal windows are useful for reducing the uncertainty of
multitaper spectral estimates, the individual spectral estimates are
not entirely independent.

\begin{figure}[t]
\centering
{\includegraphics[width=\columnwidth]{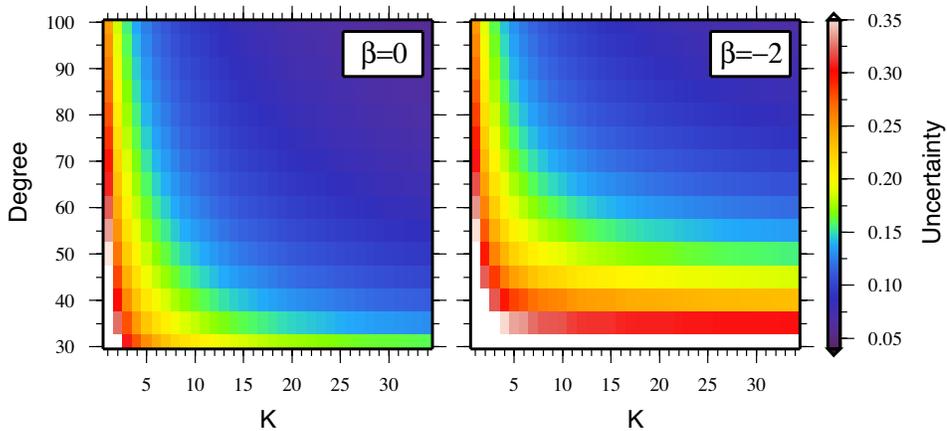}}
\caption{Optimal localized multitaper uncertainty of a white (left)
and red (right) stochastic process as a function of spherical harmonic
degree and the number of employed tapers. For the case of the red
spectrum, the uncertainty is scaled by the square root of the global
power spectrum. As in the previous figures, the windows were
constructed using $\theta_0=30^\circ$ and $L=29$. As a result of the
computationally intensive nature of these calculations, the
uncertainty was calculated only for degrees in multiples of $5$.
 \label{fig:sigwt}}
\end{figure}

The uncertainties associated with the optimal multitaper spectral
estimates are shown in the left panel of Figure~\ref{fig:sigwt} for
spherical harmonic degrees between $l=30$ and $100$ and as a function
of the number of employed localization windows. Here, we only show
results for $l > L$ since (1) the smaller degrees are influenced by
the magnitude of the degree-0 term, which is often statistically
unrelated to the other degrees for many physical processes, (2) as
shown in Figure~\ref{fig:bias}, the windowed spectrum is highly biased
for smaller degrees, and (3) it is unreasonable to expect that
wavelengths larger than the size of the window would be well resolved
in the localized spectra. As demonstrated in the previous figure, the
uncertainty of the multitaper spectral estimate decreases with
increasing number of localization windows. Furthermore, the variance
decreases with increasing spherical harmonic degree, though somewhat
more slowly. As should be readily visible, if only a few windows were
used to generate a multitaper estimate, its uncertainty could be
higher than 30\%.

\begin{figure}[t]
\centering
{\includegraphics[width=\columnwidth]{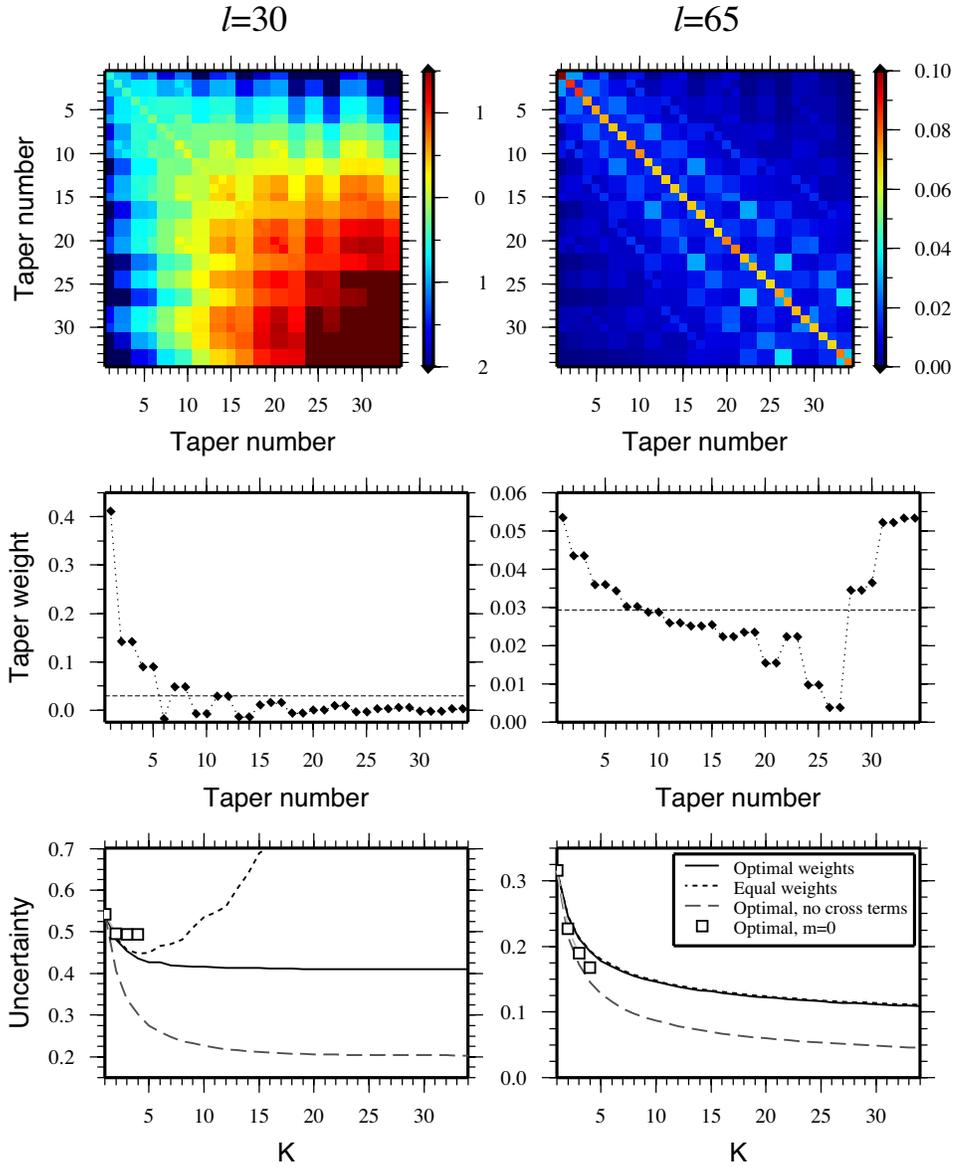}}
\caption{Same as Figure~\ref{fig:c_0_30}, but for a red global spectrum
with $\beta=-2$. The covariance matrix (top) was scaled by
$[\Sff(l)]^2$, and the uncertainty of the spectral estimates (bottom)
by $\Sff(l)$. \label{fig:c_m2_30} }
\end{figure}

Figures~\ref{fig:sigwt} (right) and~\ref{fig:c_m2_30} show analogous
results for the case when the underlying global power spectrum is red,
with $\beta=-2$. We first note that for $l=65$ in
Figure~\ref{fig:c_m2_30} (right), the covariance matrix and multitaper
uncertainties are very similar to those in the white example. As
mentioned previously, this is simply because the global power spectrum
varies slowly at high degrees and is approximately constant within the
degree range $65\pm L$. In contrast, the results for $l=30$ are
dramatically different. First, the diagonal elements of the covariance
matrix are seen to increase appreciably with increasing taper
number. Thus, the spectral estimates obtained using higher taper
numbers will possess larger uncertainties than those using smaller
taper numbers. This is a consequence of the fact that the bias for
each localization window increases significantly with increasing taper
number (see Figure~\ref{fig:bias}). The off-diagonal terms are also
seen to become increasingly important as both indices of the
covariance matrix $F_{ij}$ increase.

The lower left panel of Figure~\ref{fig:c_m2_30} shows that if one
were to use equal weights for obtaining a multitaper spectral estimate
at $l=30$, the uncertainty of this estimate would actually increase
after using more than the first five localization
windows. This is simply a result of the larger uncertainties
associated with higher taper numbers as demonstrated by the covariance
matrix. For this degree, the use of optimal weights for minimizing the
uncertainty of the multitaper estimate is critical. As shown in the
middle panel, the uncertainty of the spectral estimate is minimized in
this case by, in essence, only utilizing the first five localization
windows. The inclusion of additional windows leads to no appreciable
decrease in variance. It is further noted that given the highly
structured covariance matrix for this example, the minimum obtainable
variance is far from what would result if all the spectral estimates
were uncorrelated.

The examples shown in Figures~\ref{fig:c_0_30}--\ref{fig:c_m2_30}
demonstrate that calculating the covariance matrix $\bF$ and the
associated optimal weights is in practice only necessary for low
degrees when the underlying process is significantly red. As
computation of the covariance matrix can be somewhat time consuming
(depending on the bandwidth of the localization window, the number of
localization windows, and the spherical harmonic degree) it would be
useful to have a criterion for when the use of equal weights is
adequate, and when the use of optimal weights is necessary. The
simplest approach would be to use equal weights initially, and to then
plot the uncertainty of the multitaper spectral estimate as a function
of the number of tapers used in its construction. If the uncertainty
fails to decrease, as in the lower left panel of
Figure~\ref{fig:c_m2_30}, using optimal weights could reduce the
estimation variance significantly.

\section{Single Realizations of Stochastic Processes}
\label{section:single}

In the preceding section we assumed that the stochastic process giving
rise to the global power spectrum was known, and this allowed for the
analytic computation of the expectation~(\ref{eq:biasmt}) and
variance~(\ref{eq:mt_varfinal}) of the multitaper spectral
estimate. Unfortunately, for many physical processes, not only is the
underlying power spectrum of the stochastic process unknown, but only
a single realization is available for analysis: there is only one
gravitational field of the Earth, only one cosmic microwave
background, and so on. As Figures~\ref{fig:c_0_30}
and~\ref{fig:c_m2_30} demonstrate, the spectral estimates that result
from orthogonal localization windows are, in general, somewhat
uncorrelated. Thus, even though only a single realization of a process
might be available for analysis, each windowed spectral estimate can
be treated approximately as if it were derived from a separate
realization. As the number of localization windows used in
constructing the multitaper estimate increases, we expect the variance
of this estimate to decrease accordingly.

When the underlying global power spectrum is not known \emph{a
priori}, the primary difficulty lies in how to estimate the
uncertainty of the multitaper estimate. One approximate approach would
be to assume that the multitaper estimate is equal to the global
value, and to compute the expected variance using the expressions in
Appendix C.  If a more accurate estimate of the uncertainty were
desired, one could attempt inverting for the global power spectrum
(see below), and using this to calculate the expected multitaper
uncertainty. From the calculated covariance matrix, optimal weights
that minimize the multitaper variance could be obtained, and these
could be used to form a new multitaper estimate. By repeating this
process, one would ultimately expect to converge on the
minimum-variance multitaper spectrum estimate. This procedure,
however, suffers from having to calculate the covariance matrix $\bF$
at each degree~$l$ several times. For many problems, this approach is
unfeasible given the computationally intensive nature of the
covariance matrix calculations.

As a more practical, but necessarily approximate, approach we will
make the assumption that for a given degree~$l$ the individual
spectral estimates $\Sfifi\kth$ that contribute to the multitaper
spectrum estimate are statistically independent and Gaussian
distributed with identical variance $\sigma^2_{\Phi \Phi}$. This
amounts to assuming that the off-diagonal terms of the covariance
matrix are zero and that the diagonal terms are all equal. In this
case it is easily shown that the variance of the
estimate~(\ref{eq:mt2})--(\ref{eq:mt2a}) is

\begin{equation}
\var \! \left\{\Sfifi\mt\right\} =\sigma^2_{\Phi \Phi}\,
\sum_{k=1}^K a_k^2\quad\mbox{with}\quad
\var\!\left\{\Sfifi\kth\right\}=\sigma^2_{\Phi \Phi} 
\label{eq:varsimple}.
\end{equation}
Defining the weighted sample variance of individual
spectral estimates as 
\begin{equation}
\label{eq:s01}
\sigma^2 = \suml_{k=1}^K a_k \left( \Sfifi\kth
- \Sfifi\mt\right)^2
, 
\end{equation}
its expectation is found to be equal to
\begin{equation}
\left\langle\sigma^2\right\rangle 
=\sigma^2_{\Phi \Phi} \, \Big(1  - \sum_{k=1}^K a_k^2 \Big)
+
\sum_{k=1}^K a_k\left(
\left\langle\Sfifi\kth\right\rangle-
\left\langle\Sfifi\mt\right\rangle
\right)^2\label{kaka}
.
\end{equation}
The second term in the above equation is a measure of the variability
of the expectations of the $K$ windowed estimates $\Sfifi\kth$ about
their weighted mean (referring to Figure~\ref{fig:bias}, of the curves
of the expected values of single-taper estimates around the expected
value of the multitaper estimate). By assuming that the statistical
spread of each windowed estimate is greater than the spread of the
individual expectations, this term can be ignored.
Combining~(\ref{eq:varsimple}) and~(\ref{kaka}), the following
unbiased estimate for the variance of the multitaper estimate is
obtained:
\begin{equation}
\label{eq:var_est}
\var \left\{\Sfifi\mt\right\} \approx \sigma^2  \,
  \left(\frac{ \sum_k a_k^2}{1  - \sum_k a_k^2}\right).  
\end{equation}
In contrast to~(\ref{eq:mt_varfinal}), which requires knowledge of the
global spectrum, this estimate, though approximate, is determined
from the data alone. When the weights $a_k$ are all equal, the
variance of the multitaper spectrum estimate is simply
$\sigma^2/(K-1)$. 
Having ignored what is essentially the sample variance of the
expectations of the windowed estimates, we are insured that the
calculated uncertainty~(\ref{eq:var_est}) will be an
overestimate. However, we should also note that the assumption of each
individually tapered spectral estimate being statistically independent
will not in general be true, and this will cause the above uncertainty
to underestimate the true value. 
We note that~\cite[eq.~48]{Wieczorek+2005} previously
advocated using the sample variance $\sigma^2$ of the
individual spectral estimates as an estimate for the variance of the
multitaper estimate, which, in hindsight, is unnecessarily
conservative. 

\begin{figure}
\centering
{\includegraphics[width=\columnwidth]{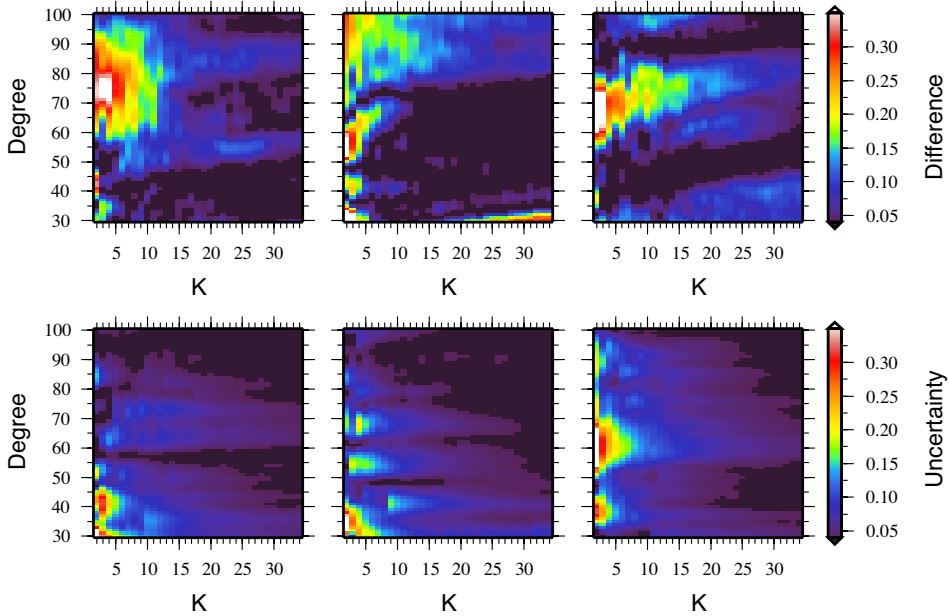}}
\caption{Difference and estimated uncertainty for three multitaper
realizations of a white stochastic process. (top) Absolute value of
the difference between the known multitaper spectrum expectation and
the single realization of the multitaper estimate. (bottom) The
estimated uncertainty of the multitaper spectrum estimate calculated
from~(\ref{eq:var_est}) using equal weights.  Between 2 and 34 tapers
were used in generating the multitaper estimates, the tapers were
constructed using $\theta_0=30^\circ$ and $L=29$, and the data were
localized at the North pole. For comparison purposes, the color bars
possess the same range as those in Figure~\ref{fig:sigwt}.
\label{fig:single} }
\end{figure}

In order to assess the suitability of the above approximations to
estimate the uncertainty via~(\ref{eq:var_est}), we generated three
realizations of a white stochastic process and calculated localized
multitaper estimates. In the upper panel of Figure~\ref{fig:single},
the absolute value of the difference between the multitaper estimate
and its expectation based on the known input spectrum is shown as a
function of degree~$l$ and number of tapers $K$ for the three
realizations. As is readily seen, the difference between the two
almost everywhere decreases with increasing number of employed
localization windows. Furthermore, comparison with
Figure~\ref{fig:sigwt} shows that the difference between the two is
compatible with the expected uncertainty of the multitaper
estimate. The bottom panel plots the uncertainty of the multitaper
estimate using~(\ref{eq:var_est}), and this is seen to be generally
comparable to the difference between the multitaper expectation and
single realization in the upper panel. Nevertheless, it should be
noted that this estimation of the uncertainty underestimates the true
value as shown in Figure~\ref{fig:sigwt} by a small factor. This is
most likely a result of the fact that the individual spectral
estimates are not uncorrelated as we assumed in
deriving~(\ref{eq:var_est}).

Finally, we note that it is possible under certain circumstances to
invert for the global power spectrum using~(\ref{eq:biasmt}) combined
with knowledge of the multitaper spectral estimates and their
uncertainties. As shown in~\cite{Wieczorek+2005}, this equation can be
written in matrix notation as
\begin{equation}
\label{eq:m_smooth}
\left\langle \bSfifi \right\rangle = 
\bM\mt \,
\bSff ,
\end{equation} 
where $\bSfifi$ is a vector containing
the $L_{\Phi \Phi}+1$ multitaper spectral estimates,
$\bSff$ is a vector containing the $L_{\Phi \Phi}+L+1$
elements of the global power spectrum, and $\bM\mt$ is an
$(L_{\Phi \Phi}+1) \times (L_{\Phi \Phi}+L+1)$ matrix that maps the
latter into the former. Assigning  the index $0$ to the first row and
column of $\bM$, the elements are given by
\begin{equation}
\label{mat1}
M\mt_{ij} =  \sum_{l=0}^L \sum_{k=1}^K a_k \, \Shh\kth(l)   \left( 
C_{l0j0}^{i0}
\right) ^2
.
\end{equation} 
The individual linear equations of~(\ref{eq:m_smooth}) could be
weighted by the measurement uncertainties by dividing each row of
$\bSfifi$ and $\bM\mt$ by the multitaper standard deviation obtained
from~(\ref{eq:var_est}).

Inverting $\bM\mt$ to obtain $\bSff$ from $\bSfifi$ is an
underdetermined inversion problem, as the dimension of $\bSff$ will
always be greater than that of $\bSfifi$, and $\bM\mt$ is never
full-rank. Hence, most of our attempts to invert~(\ref{eq:m_smooth})
for the global power spectrum, using a variety of linear techniques,
have been unsatisfactory for this example. For instance, choosing that
solution for which the norm of $\Sff$ is minimized yielded a power
spectrum with both positive and negative values, a situation that is
clearly unphysical. Truncating the matrix $\bM$ to be square yielded
similar results. A non-negative least squares inversion
\cite[Chap.~23]{Lawson+95} yielded a solution for which the majority
of the $\Sff(l)$ were zero. One method that did yield acceptable
results was to assume that $\Sff$ was constant in bins of width
$\Delta l$. However, for the problem at hand, $\Delta l$ was required
to be greater than about 15 in order to obtain positive values with
reasonable variances. More sophisticated non-linear and Monte Carlo
techniques that utilize positivity constraints and upper and lower
bounds are worth investigating. Alternatively, one could parametrize
the global power spectrum by a smoothly varying function (such as a
power law) and invert for the parameter values that best fit the
observed multitaper spectral estimates. A maximum-likelihood inversion
approach is described in~\cite{Dahlen+2008} and~\cite{Hansen+2002}.

\section{Concluding Remarks}

In this study, we have demonstrated how the global power spectrum of a
stochastic process on the sphere can be estimated from spatially
limited observation domains. In particular, the act of restriction to
a certain region can be formulated as a windowing operation, and the
expectation of the windowed power spectrum has been shown to be
related to the global power spectrum by a convolution-type
equation. The best localization windows are those that are spatially
concentrated in the region of interest, and yet have as small an
effective spectral bandwidth as possible. Solution of a simple
optimization problem yields a family of orthogonal functions, and a
multitaper spectrum estimate can be constructed using those windows
with good spatiospectral localization properties. The multitaper
estimate has the benefits that the data are more evenly weighted and
that it has a reduced variance when compared to single-window
estimates. It is straightforward to choose the weights used in
constructing the multitaper estimate in order to minimize the
estimation variance.

While much of the theoretical groundwork has been developed for the
problem of localized spectral estimation on the sphere, several
promising lines of future research could improve upon the procedure
developed in this work. In particular, we have only concerned
ourselves with localization windows that are solutions to the
spherical-cap concentration problem
\cite{Grunbaum+82,Simons+2006,Wieczorek+2005}. While sufficient for
many problems, such as estimating the localized spectral properties of
planetary gravitational fields and topography, more complex localization
windows might be desired for other applications. The
techniques developed here are easily generalizable to other
concentration domains~\cite{Simons+2006b,Simons+2006}, and we do not
expect the general character of our results to differ significantly
from those presented here.

As a second line of future research, we note that we have restricted
ourselves to using only a single localization window when calculating
the localized cross-power spectrum $\Sfiga\kth$. However, as
demonstrated in Appendix~B, we could have localized each field by a
different window to form localized cross-power spectra
$\Sfiga^{(j,k)}$. Using two different windows, up to $K(K+1)/2$
individual cross-power spectral estimates are, in principle, available
for analysis compared to $K$ as used in this study. The covariance
properties of these windows remain to be investigated, as well as
whether their use would significantly decrease the multitaper
estimation variance.

Of more practical concern are the computational demands to obtain the
covariance matrix.  Generating the left and right panels of
Figure~\ref{fig:c_0_30} took about 12 hours and 2 days of dedicated
time on a modern desktop computer, respectively. The calculations for
Figure~\ref{fig:sigwt} took considerably longer, about one
month. Clearly, such computations will become increasingly infeasible
as the spherical harmonic degree increases, and alternative means will
eventually be necessary for computing the covariance matrix, optimal
weights, and uncertainties. Asymptotic relations for the
Clebsch-Gordan coefficients and/or covariance matrix may be used to
simplify these numerical computations at high
degrees~\cite{Dahlen+2008}.

Finally, we note that most of our discussion has emphasized the
quantification of the bias and uncertainty of the multitaper spectrum
estimate. However, in practice, one is often instead interested in
obtaining an unbiased estimate of the global power spectrum. While one
can in principle invert~(\ref{eq:biasmt}) for this quantity, the
standard linear inversion techniques discussed in
Section~\ref{section:single} yielded mixed results. More sophisticated
non-linear and Monte Carlo techniques utilizing bounds and positivity
constraints may be worth investigating in this context.

\appendix

\section{Real and Complex Spherical Harmonics}

The derivations in this study are considerably simplified if complex
spherical harmonics are used. By inserting the identities
\begin{equation}
\cos{\beta} = \frac{e^{i\beta} + e^{-i\beta}}{2} \quad \mbox{and}
\quad \sin{\beta} = \frac{e^{i\beta} - e^{-i\beta}}{2i} 
\end{equation} 
into~(\ref{eq:sh_ex1}), the spherical harmonic expansion of
a real function $f$ is expressed as 
\begin{eqnarray}
f(\Omega) &=& \sum_{l=0}^{\infty} \sum_{m=-l}^{l} f_{lm}
Y_{lm}(\Omega)\nonumber\\  
&=& \sum_{l=0}^{\infty} \sum_{m=0}^{l} \left[ f_{lm}
  \left(\frac{e^{im\phi} + e^{-im\phi}}{2}\right) +
  f_{l-m}\left(\frac{e^{im\phi} - e^{-im\phi}}{2i} \right) \right]
\bar{P}_{lm}(\cos \theta)\nonumber\\  
&=&  \sum_{l=0}^{\infty} \sum_{m=0}^{l}
\left[e^{im\phi}\left(\frac{f_{lm} - if_{l-m}}{2}\right) +
  e^{-im\phi}\left(\frac{f_{lm} +
    if_{l-m}}{2}\right)\right]\bar{P}_{lm}(\cos \theta)\nonumber\\ 
&=&\sum_{l=0}^{\infty} \sum_{m=-l}^{l} f_l^m\, Y_l^m(\Omega),
\end{eqnarray}
 where the complex spherical harmonics, $Y_l^m$, are defined as
\begin{equation}
Y_l^m(\Omega) =  \sqrt{ 2l+1}
\sqrt{\frac{(l-m)!}{(l+m)!}}\, P_{lm}(\cos\theta) \, e^{im\phi} .
\end{equation}
Since 
\begin{equation}
P_{l\,-m}=(-1)^m\frac{(l-m)!}{(l+m)!}\,P_{lm},
\end{equation} 
the complex harmonics satisfy the identity  
\begin{equation}
{Y_l^m}^*(\Omega) = (-1)^m \, Y_l^{-m}(\Omega),
\label{eq:ident1}
\end{equation}
and normalization
\begin{equation}
\label{eq:csh_orthog}
\frac{1}{4\pi}\int_\Omega {Y_{l}^{m}}^*(\Omega) Y_{l'}^{m'}(\Omega)
\domg =   
\delta_{ll'}\, \delta_{mm'}
,
\end{equation}
where the superscript $^*$ indicates complex conjugation. To
be consistent with the definition of the real spherical harmonic
functions in Section 2.1, the complex harmonics here also do not include
the Condon-Shortley phase factor of $(-1)^m$ that generally appears in
the physics and seismology communities. Regardless, we note that the
inclusion or exclusion of this phase will not affect the
results presented in this paper.
The complex coefficients are related to the real coefficients by
\begin{equation}
\label{eq:convert}
f_l^m =  \left \lbrace \begin{array}{ll} (f_{lm} - if_{l-m}) /
	\sqrt{2} & \mbox{if $m >0$} \\ 
	f_{l0} & \mbox{if $m = 0$}\\
	 (-1)^m \, f_l^{-m*} & \mbox{if $m <0$},
	\end{array} \right.
\end{equation}
and by using the orthogonality properties of the spherical
harmonics, these can be shown to be related to the real function $f$
by the relation 
\begin{equation} 
\label{eq:flm}
f_{l}^{m} = \frac{1}{4\pi} \int_\Omega f(\Omega)\,
{Y_{l}^{m}}^*(\Omega) \domg
.
\end{equation}
It is straightforward to show that the total power of a real function
$f$ is related to its complex spectral coefficients by a
generalization of Parseval's theorem: 
\begin{equation}
\label{eq:Apower}
\frac{1}{4\pi} \int_\Omega \left[ f(\Omega)\right]^2\domg =
\sum_{l=0}^\infty \Sff(l) 
,
\end{equation}
where the power spectrum is
\begin{equation}
\Sff(l)  = \sum_{m=-l}^l f_l^{m} {f_l^{m}}^* 
.
\end{equation}
Similarly, the cross-power of two real functions $f$ and $g$ is given
by 
\begin{equation}
\label{eq:cross_power}
\frac{1}{4\pi} \int_\Omega f(\Omega) \,g(\Omega) \domg =
\sum_{l=0}^{\infty} \Sfg(l) , 
\end{equation}
where the cross-power spectrum is
\begin{equation}\label{eq:cross_power2}
\Sfg(l) = \sum_{m=-l}^l f_l^{m} {g_l^{m}}^*
.
\end{equation}
If the functions $f$ and $g$ have a zero mean (i.e., their degree-0
terms are equal to zero), then $\Sff(l)$ and $\Sfg(l)$ represent the
contributions to their variance and covariance, respectively, for
degree~$l$.

\section{Bias of a Localized Spectral Estimate}

In this section, expressions will be derived that relate the
cross-power spectrum of two global fields to their windowed
equivalents. We assume that the real spherical harmonic
coefficients of a function $f$ are zero-mean random variables, and
that the power spectrum of the function is isotropic, i.e. depends
only upon degree~$l$: 
\begin{equation}
\langle f_{lm} f_{l'm'} \rangle = \frac{\Sff(l)}{(2l+1)}\, \delta_{ll'}
\, \delta_{mm'}, \label{eq:b1}
\end{equation} 
where $\langle \cdots \rangle$ signifies the expectation
operator.  When considering two fields, it will be assumed in 
a similar manner that their cross-power at a given degree is 
also isotropic:
\begin{equation}
\langle f_{lm} g_{l'm'} \rangle = \frac{\Sfg(l)}{(2l+1)}
\,\delta_{ll'} \, \delta_{mm'}. 
\end{equation} 
It can be verified using~(\ref{eq:convert}) that 
\begin{equation}
\langle f_{lm} f_{l'm'} \rangle = \langle {f_l^m}^* f_{l'}^{m'}
\rangle = \langle {f_l^m}^* f_{l'}^{m'} \rangle^* , 
\end{equation}
and
\begin{equation}
\langle f_{lm} g_{l'm'} \rangle = \langle {f_l^m}^* g_{l'}^{m'}
\rangle = \langle {f_l^m}^* g_{l'}^{m'} \rangle^*. 
\end{equation}

The goal of this appendix is to find an expression for the expectation
of the cross-power spectrum of the functions $f$ and $g$, each
localized by a different data taper $h^{(i)}$ and $h^{(j)}$,
respectively. This quantity will be denoted as
$\langle\Sfiga^{(i,j)}(l)\rangle$, where $\Phi$ and $\Gamma$ represent
the localized fields $f \,h^{(i)}$ and $g \,h^{(j)}$,
respectively. The spectral bandwidth of each localization window will be
assumed to be the same. We start with the product of two
windowed coefficients for a given degree and order:
\begin{eqnarray}
\Phi_l^{m(i)} \, \Gamma_l^{m(j)*} &=&\frac{1}{4\pi} \int_{\Omega}
    {\big [}h^{(i)}(\Omega) 
\,f(\Omega) {\big ]} Y_l^{m*}(\Omega) \domg \nonumber \\ 
&& \times
\frac{1}{4\pi} \int_{\Omega'} {\big [} h^{(j)*}(\Omega')
    \,g^*(\Omega') {\big ]} 
Y_l^m(\Omega') \domg'.
\end{eqnarray}
Expanding the windowed functions in spherical harmonics, and utilizing
the short-hand notation 
\begin{equation}
\suml_{lm}^{L} = \suml_{l=0}^{L} 
\suml_{m=-l}^{l} \quad \mbox{and} \quad \suml_{l}^{L} =
\suml_{l=0}^{L},
\end{equation}
yields
\begin{eqnarray} 
\Phi_l^{m(i)} \,\Gamma_l^{m(j)*} &=& \frac{1}{(4\pi)^2}
 \suml_{l_1m_1}^{\Lh} h_{l_1}^{m_1(i)}
 \suml_{l_2m_2}^{\infty} f_{l_2}^{m_2}
 \suml_{l_3m_3}^{\Lh} 
  h_{l_3}^{m_3(j) *} \suml_{l_4m_4}^{\infty} 
  g_{l_4}^{m_4*}  \nonumber \\ 
&&\times \int_{\Omega} Y_{l_1}^{m_1} \, Y_{l_2}^{m_2}\,  Y_l^{m*} \,
 d\Omega \int_{\Omega'}  Y_{l_3}^{m_3*}\, Y_{l_4}^{m_4*}\,   Y_l^m \,
 d\Omega'.  
\label{eq:a}
\end{eqnarray}
Averaging this equation over all possible combinations of
the random variables gives its expectation:
\begin{eqnarray} 
\left\langle\Phi_l^{m(i)}\, \Gamma_l^{m(j)*}\right\rangle &=&
 \frac{1}{(4\pi)^2} 
 \suml_{l_1m_1}^{\Lh} h_{l_1}^{m_1(i) }
 \suml_{l_3m_3}^{\Lh}
 h_{l_3}^{m_3(j)*}\suml_{l_2m_2}^{\infty}
 \frac{\Sfg(l_2)}{(2l_2+1)} \label{eq:b}  \\  
&&\times \int_{\Omega} Y_{l_1}^{m_1} \,Y_{l_2}^{m_2}\,   {Y_l^m}^* \,
 d\Omega 
  \int_{\Omega'} Y_{l_3}^{m_3*}\,{Y_{l_2}^{m_2}}^*\,  Y_l^m \,
 d\Omega'.  \nonumber 
\end{eqnarray} 
The integral of a triple product of spherical harmonics is real and
can be evaluated by a well-known relationship 
involving Clebsch-Gordan coefficients~\cite[p.~148]{Varshalovich+88}
\begin{equation}
\int_{\Omega} {Y_{l_1}^{m_1}}\,  Y_{l_2}^{m_2}\,  Y_l^{m*} \domg
=  4 \pi \sqrt{\frac{(2l_1+1)(2l_2+1)}{(2l+1)}} C_{l_10l_20}^{l0} \,
C_{l_1m_1l_2m_2}^{lm}, 
\label{eq:cg}
\end{equation}
which is non-zero only when the following selection rules are
satisfied~\cite{Varshalovich+88}: 
\begin{eqnarray}
&&m = m_1 + m_2 \label{eq:s1} \\ 
&&|m|\le l;\, |m_1| \le l_1; \, |m_2|\le l_2 \label{eq:s2}\\ 
&&|l_1-l_2| \le l \le l_1 + l_2 \label{eq:s3}\\
&&|l_2-l| \le l_1 \le l_2 + l \label{eq:s5}\\
&&|l-l_1| \le l_2 \le l + l_1  \label{eq:s4}\\
&&l_1 + l_2 + l = \mathrm{even}.  \label{eq:s6}
\end{eqnarray}
Combining~(\ref{eq:b}) and~(\ref{eq:cg}), and making use
of~(\ref{eq:ident1}), yields 
\begin{eqnarray} 
\left\langle\Phi_l^{m(i)}\, \Gamma_l^{m(j)*}\right\rangle& =&
 \suml_{l_1m_1}^{\Lh} h_{l_1}^{m_1(i) } \suml_{l_3m_3}^{\Lh}
  h_{l_3}^{m_3(j)*}\suml_{l_2m_2}^{\infty} \Sfg(l_2)   \\ 
&&\times \, \frac{\sqrt{(2l_1+1)(2l_3+1)}}{2l+1} \, C_{l_10l_20}^{l0}
 \, C_{l_30l_20}^{l0} \,  C_{l_1m_1l_2m_2}^{lm} \,
 C_{l_3-m_3l_2-m_2}^{l-m},  \nonumber  
\label{eq:c}
\end{eqnarray}
where a phase factor is set equal to unity because
of~(\ref{eq:s1}). We next sum this entire equation over all values of 
$m$, employ the symmetry relationship of the Clebsch-Gordan
coefficients~\cite[p.~245]{Varshalovich+88}
\begin{equation}
C_{l_1-m_1l_2-m_2}^{l-m} = (-1)^{l_1+l_2-l} C_{l_1m_1l_2m_2}^{lm},
\end{equation}
set a phase factor equal to unity because of~(\ref{eq:s6}),
and rearrange the sum over $m_2$ to obtain 
\begin{eqnarray} 
\left\langle \Sfiga^{(i,j)}(l) \right\rangle &=& 
 \suml_{l_1m_1}^{\Lh} h_{l_1}^{m_1(i) } \suml_{l_3m_3}^{\Lh}
  h_{l_3}^{m_3(j)*}\suml_{l_2}^{\infty} \Sfg(l_2) \\ 
&&{\hspace{-3em}}\times  \frac{\sqrt{(2l_1+1)(2l_3+1)}}{2l+1}
 C_{l_10l_20}^{l0} \, 
 C_{l_30l_20}^{l0}
 \suml_{m=-l}^{l}\suml_{m_2=-l_2}^{l_2}C_{l_1m_1l_2m_2}^{lm} 
 \, C_{l_3m_3l_2m_2}^{lm}. \nonumber
\label{eq:d}
\end{eqnarray}
The final sum over $m$ and $m_2$ is greatly simplified by
use of the identity~\cite[p.~259]{Varshalovich+88} 
\begin{equation}
\sum_{\alpha}\sum_{\gamma} C_{a \alpha b \beta}^{c  
  \gamma} \, C_{a \alpha b' 
\beta'}^{c \gamma} = \frac{(2c+1)}{(2b+1)} \delta_{bb'}\,
\delta_{\beta \beta'} \label{eq:cg_identity}
,
\end{equation}
where the summations are implicitly over all values which
are non-zero, and the symmetry
relationship~\cite[p.~245]{Varshalovich+88} 
\begin{equation}
C_{l_1m_1l_2m_2}^{lm} = (-1)^{l_1+l_2-l}C_{l_2m_2l_1m_1}^{lm}.
\end{equation}
Taking into account the selection rule~(\ref{eq:s4}),
(\ref{eq:d}) can be succinctly written as
\begin{equation} 
\left\langle \Sfiga^{(i,j)}(l) \right\rangle = 
 \suml_{l_1m_1}^{\Lh} h_{l_1}^{m_1(i)}
  h_{l_1}^{m_1(j)*}\suml_{l_2=|l-l_1|}^{l+l_1} \Sfg(l_2)
 \left(C_{l_10l_20}^{l0}\right)^2, 
\end{equation}
where it is clear that this quantity is real. While the sum
over $m_1$ might sometimes be zero, in general, this will not be the
case, even for when $h^{(i)}$ and $h^{(j)}$ are orthogonal. Thus, if
$K$ tapers are being employed in the spectral estimation procedure, up
to $K(K+1)/2$ cross-spectral estimates can be obtained. For the
specific case where $f$ and $g$ are localized by the same window, the
expectation of the localized cross-power spectrum is, after a change
of variables,
\begin{equation}
\left\langle \Sfiga(l) \right\rangle = 
 \suml_{j=0}^{\Lh} \Shh(j)
 \suml_{i=|l-j|}^{l+j} \Sfg(i)  \left(C_{j0i0}^{l0}\right)^2.
 \label{eq:Abias}
\end{equation} 
For computational purposes, we note that the Wigner 3-$j$
symbols are related to the Clebsch-Gordan coefficients by the
definition~\cite[p.~236]{Varshalovich+88}
\begin{equation}
C_{l_1m_1l_2m_2}^{lm} = (-1)^{l_1-l_2+m}\sqrt{2l+1}
\left( 
\begin{array}{ccc}
l_1 & l_2 & l \\ m_1 & m_2 & -m
\end{array}
\right)
. 
\end{equation} 
Algorithms for calculating the Wigner 3-$j$ symbols are
discussed by~\cite{Luscombe+98} and~\cite{Schulten+75}.

\section{Variance of a Localized Spectral Estimate}

A multitaper estimate for the localized cross-power spectrum of two
fields $f$ and $g$ will be defined as a weighted average of direct
spectral estimates obtained using $K$ orthogonal tapers:
\begin{equation}
S\mt_{\Phi \Gamma}(l) = \sum_{i=1}^K \sum_{j=1}^K a_{ij} \,
S^{(i,j)}_{\Phi \Gamma}(l)
,
\label{eq:mt1}
\end{equation}
where the elements of $a_{ij}$ are the weights applied to
the cross-spectral estimate subject to the constraint  
\begin{equation}
\suml_{i=1}^K \suml_{j=1}^K a_{ij}=1. 
\end{equation} 
It is easy to verify that this quantity is real when
$a_{ij}$ is symmetric. To simplify the following derivations, and to
reduce the amount of time required to calculate these numerically, we
will consider only the case where $i=j$. Further research is required
to determine the benefit of employing a spectral estimate obtained
using two different tapers. In this case, the multitaper
spectral estimate reduces to
\begin{equation}
S\mt_{\Phi \Gamma}(l) = \sum_{k=1}^K a_k \,
S\kth_{\Phi \Gamma}(l)
\label{eq:Amt2}
,
\end{equation}
with the constraint 
\begin{equation}
\suml_{k=1}^K a_{k}=1. 
\label{eq:Amt2a}
\end{equation} 
This appendix seeks to determine the variance of such a spectral estimate. 

We start with the definitions of variance and covariance of
complex variables: 
\begin{eqnarray}
\label{eq:var1}
\var \left\{ \sum_{i=1}^N a_i \, X_i \right\} &=&
\sum_{j=1}^N \sum_{k=1}^N a_j \,  \cov\{X_j , X_k\} \, a_k \\ 
\label{eq:covar}
\cov\{X_j , X_k\} &=& \langle X_j X^*_k \rangle -  \langle X_j
\rangle  \langle X^*_k \rangle. 
\end{eqnarray}
which give the following expression for the variance
of~(\ref{eq:Amt2}): 
\begin{equation}
\label{eq:mt_var1}
\var \left\{S\mt_{\Phi \Gamma}(l)
\right\}= \sum_{j=1}^K \sum_{k=1}^K a_j \,
\cov \left\{S^{(j)}_{\Phi \Gamma}(l) , S\kth_{\Phi
\Gamma}(l)\right\} \, a_k
.
\end{equation} 
By using the definition of the cross-power
spectrum~(\ref{eq:cross_power})--(\ref{eq:cross_power2}) with the
identity   
\begin{equation}
\cov \left\{ \sum_{i=1}^N X_i, \sum_{j=1}^M X_j
\right\} = \sum_{i=1}^N \sum_{j=1}^M \cov\{X_i, X_j\},
\end{equation}
the covariance of two spectral estimates using tapers $j$ and $k$ can
be written as 
\begin{equation}
\label{eq:inter}
\cov \left\{S^{(j)}_{\Phi \Gamma}(l),
S\kth_{\Phi \Gamma}(l)\right\} = \sum_{m=-l}^l
\sum_{m'=-l}^l \cov\left\{ \Phi_{l}^{m(j)}
\Gamma_{l}^{m(j)*}, \Phi_{l}^{m'(k)} \Gamma_{l}^{m'(k)*} \right\}.
\end{equation}
We proceed by using Isserlis' theorem~\cite{Walden+94}
\begin{equation} 
\label{eq:isserlis}
\cov\{Z_1 \, Z_2, Z_3 \, Z_4\} = \cov\{Z_1,
Z_3\} \, \cov\{Z_2, Z_4\}
+\,  \cov\{Z_1, Z_4\} \, \cov\{Z_2,
Z_3\},
\end{equation} 
which is valid for zero-mean Gaussian complex random variables
$Z_i$. Given that we have previously assumed that the coefficients
$f_{lm}$ and $g_{lm}$ have a zero mean, so will the spectral
coefficients of the localized fields $\Phi_{lm}$ and $\Gamma_{lm}$. We
may then rely on a central-limit theorem~\cite{Wieczorek+2005} to
subsequently assume that the localized coefficients will approach a
Gaussian distribution. Alternatively, if we assume that the
coefficients of the unwindowed fields were Gaussian to begin with,
then it is easily shown that the windowed coefficients will be Gaussian as
well. Under these conditions (\ref{eq:inter}) can be written as
\begin{eqnarray}
\cov\left\{S^{(j)}_{\Phi \Gamma}(l) , S\kth_{\Phi
\Gamma}(l)\right\}\hspace{-0.5em}&=&\hspace{-0.5em}\sum_{m=-l}^l \sum_{m'=-l}^l
\left(\cov\left\{\Phi_{l}^{m(j)}, \Phi_{l}^{m'(k)}\right\} \,
\cov\left\{\Gamma_{l}^{m(j)*}, \Gamma_{l}^{m'(k)*}\right\}
\right. \nonumber \\ 
&&+
\left. \cov\left\{ \Phi_{l}^{m(j)}, \Gamma_{l}^{m'(k)*}
\right\} \, \cov\left\{ \Gamma_{l}^{m(j)*}, \Phi_{l}^{m'(k)}
\right\} \right),\label{doh}
\end{eqnarray}
and it is trivial to generalize~(\ref{eq:c}) to show that
\begin{eqnarray}
\label{eq:f}
\cov\left\{ \Phi_{l}^{m(j)}, \Gamma_{l}^{m'(k)}
\right\} = \left\langle\Phi_l^{m(j)}\, \Gamma_l^{m'(k)*}\right\rangle
\hspace{-0.5em}&=&\hspace{-0.5em}
\suml_{l_1m_1}^{\Lh} h_{l_1}^{m_1(j) } \suml_{l_3m_3}^{\Lh}
  h_{l_3}^{m_3(k)*}\suml_{l_2m_2}^{\infty}
  \Sfg(l_2)\nonumber
\\
&&\hspace{-15em}\times  \, \frac{\sqrt{(2l_1+1)(2l_3+1)}}{2l+1}
C_{l_10l_20}^{l0} \, C_{l_30l_20}^{l0} \, C_{l_1m_1l_2m_2}^{lm} \, 
C_{l_3m_3l_2m_2}^{lm'}.   
\end{eqnarray}
For computational purposes, we note that the sums over $l_2$ and $m_2$
are considerably restricted as a result of the selection
rules~(\ref{eq:s1})--(\ref{eq:s6}), which imply
\begin{eqnarray}
&&m_2 = m-m_1,\\
&&m_2 = m' - m_3,\\
&&m-m_1 = m' - m_3, \label{eq:m13}\\
&&l_1 + l_2 + l = \mathrm{even},\\
&&l_3 + l_2 + l = \mathrm{even},\\
&&l_1 + l_3  = \mathrm{even}.
\end{eqnarray}
The expression for the variance is somewhat simplified when
only one field, $f$, is considered. Noting that 
\begin{eqnarray}
\left\langle\Phi_l^{m(j)*}\, \Phi_l^{m'(k)*}\right\rangle &=&
\left\langle\Phi_l^{m(j)}\, \Phi_l^{m'(k)}\right\rangle^*,\\ 
\left\langle\Phi_l^{m(j)*}\, \Phi_l^{m'(k)}\right\rangle &=&
\left\langle\Phi_l^{m(j)}\, \Phi_l^{m'(k)*}\right\rangle^*, 
\end{eqnarray}
the variance of the multitaper spectral estimate can be
written as 
\begin{equation}
\label{eq:Amt_varfinal}
\var \left\{S\mt_{\Phi \Phi}(l)
\right\}=  \sum_{j=1}^K \sum_{k=1}^K a_j \,  F_{jk} \, a_k ,
\end{equation}
where $\bF$ implicitly depends upon $l$ and $\Sff$ and is given
by 
\begin{equation}
F_{jk} = \suml_{m=-l}^l\suml_{m'=-l}^l
\left( \left|\langle \Phi_l^{m(j)} \Phi_l^{m'(k)*}\rangle \right|^2 +
\left|\langle \Phi_l^{m(j)} \Phi_l^{m'(k)} \rangle \right|^2
\right). \label{eq:fjk}
\end{equation}
Using the negative angular order symmetry relations~(\ref{eq:convert})
combined with the fact that the above sums are performed over all
values of $m$, $\bF$  further simplifies to
\begin{equation}
F_{jk} = 2 \suml_{m=-l}^l\suml_{m'=-l}^l
 \left|\langle \Phi_l^{m(j)} \Phi_l^{m'(k)*}\rangle \right|^2 .
\label{eq:fjk2}
\end{equation}
The covariance matrix is naturally symmetric.

Finally, we note that~(\ref{eq:f}), and hence the computation of
$\bF$, can be simplified for the case where the windows are solutions
of the spherical-cap concentration problem~\cite{Simons+2006}. For
this situation, each window $j$ and $k$ has non-zero \emph{real}
spherical harmonic coefficients only for a single angular order $m_j$
and $m_k$, respectively. When the windows are expressed in complex
form, this implies that the only non-zero coefficients are for
$m_1=\pm m_j$, $m_3 = \pm m_k$, $l_1 \ge | m_j |$ and $l_3 \ge | m_k|
$.

\section{Correlation of Multitaper Spectral Estimates}

It was shown in Appendix B and (\ref{eq:biasmt}) that the expectation
of a multitaper spectral estimate at degree~$l$ depends upon the
global power spectrum within the degree range $l\pm L$,
where $L$ is the bandwidth of the localization windows. 
It is thus natural to expect that multitaper spectral estimates
separated by less than $2L$ degrees will be partially
correlated. Following the methodology presented in Appendix C, this
correlation is quantified by the covariance of the two multitaper
spectral estimates, which can be shown to equal
\begin{equation}
\label{eq:covarmt1}
\mathrm{cov} \left\{S^{(mt)}_{\Phi \Gamma}(l), S^{(mt)}_{\Phi
\Gamma}(l')\right\} = \sum_{i=1}^K \sum_{j=1}^K a_i \,F_{ij}^{ll'} \, a_j,
\end{equation}
where
\begin{eqnarray}
F_{ij}^{ll'} &= & \sum_{m=-l}^l \sum_{m'=-l'}^{l'}\label{doho} 
\left(\mathrm{cov}\left\{\Phi_{l}^{m(i)}, \Phi_{l'}^{m'(j)}\right\} \,
\mathrm{cov}\left\{\Gamma_{l}^{m(i)*}, \Gamma_{l'}^{m'(j)*}\right\}
\right.  \\ 
&&{\hspace{4.5em}}+
\left. \mathrm{cov}\left\{ \Phi_{l}^{m(i)}, \Gamma_{l'}^{m'(j)*}
\right\} \, \mathrm{cov}\left\{ \Gamma_{l}^{m(i)*}, \Phi_{l'}^{m'(j)}
\right\} \right),\nonumber
\end{eqnarray}
\noindent and
\begin{eqnarray}
\label{eq:f2}
\mathrm{cov}\left\{ \Phi_{l}^{m(i)}, \Gamma_{l'}^{m'(j)} \right\} =
\left\langle\Phi_l^{m(i)}\, \Gamma_{l'}^{m'(j)*}\right\rangle
\hspace{-0.5em}&=&\hspace{-0.5em} 
\sum\limits_{l_1m_1}^{\Lh} h_{l_1}^{m_1(i) }
\sum\limits_{l_3m_3}^{\Lh}
h_{l_3}^{m_3(j)*}\sum\limits_{l_2m_2}^{\infty} S_{fg}(l_2) \nonumber 
\\ &&
\hspace{-15em}\times \, \sqrt{\frac{(2l_1+1)(2l_3+1)}{(2l+1)(2l'+1)}}
C_{l_10l_20}^{l0} \, C_{l_30l_20}^{l'0} \, 
C_{l_1m_1l_2m_2}^{lm} \, C_{l_3m_3l_2m_2}^{l'm'}.
\end{eqnarray}
If only cross-power spectra of a single function are being considered,
the matrix~$\bF$ can be considerably simplified to
\begin{equation}
F_{ij}^{ll'} = 2 \sum\limits_{m=-l}^l\sum\limits_{m'=-l'}^{l'}
 \left|\langle \Phi_l^{m(i)} \Phi_{l'}^{m'(j)*}\rangle \right|^2 .
\label{eq:fikll}
\end{equation}

\section*{Acknowledgments}
We thank two anonymous reviewers for comments that helped clarify
portions of this manuscript. Financial support for this work has been
provided by the U.~S.~National Science Foundation under Grant
EAR-0710860 awarded to FJS at Princeton University, and by
U.~K.~Natural Environmental Research Council New Investigator Award
NE/D521449/1 and Nuffield Foundation Grant for Newly Appointed
Lecturers NAL/01087/G to FJS at University College London. Software
for performing the computations in this paper can be found on the
authors' web sites. This is IPGP contribution 2236.

{\footnotesize
\centerline{\rule{9pc}{.01in}}
\bigskip
\centerline{Received September 30, 2006}
\medskip
\centerline{Revision received \today} 
\medskip
\centerline{Equipe d'Etudes Spatiales et Plan{\'e}tologie}
\centerline{Institut de Physique du Globe de Paris} 
\centerline{94107 Saint Maur, France}
\centerline{e-mail: wieczor@ipgp.jussieu.fr}
\medskip
\centerline{Department of Earth Sciences, University College of London} 
\centerline{Gower Street, London, WC1E 6BT, United Kingdom}
\centerline{\it and} 
\centerline{Department of Geosciences, Princeton University} 
\centerline{Guyot Hall, Princeton, NJ 08544, USA}
\centerline{e-mail: fjsimons@alum.mit.edu}
}
\end{document}